**Title:** Our Solar System Neighborhood: Three Diverging Tales of Planetary Habitability and Windows to Earth's Past and Future

**Authors:** Stephen R. Kane, Richard Ernst, Cedric Gillmann, Christopher Jones, Timothy Lyons, Christopher Tino

**Abstract**

Understanding planetary habitability is one of the major challenges of the current scientific era, particularly given the discovery of a large and diverse terrestrial exoplanet population. Discerning the primary factors that contribute to planetary habitability may be extracted through a detailed examination of the terrestrial planets within the Solar System, most particularly Venus, Earth, and Mars, and the evolution of their interiors and atmospheres through time. Here, we provide a detailed description of the fundamental properties of these three planets, the effects of solar evolution, and the potential contributions of these various aspects toward driving their evolutionary pathways. We argue that evolution of Venus, Earth, and Mars provide essential templates from which a more comprehensive approach toward the study of planetary habitability may be derived.

**Section 1. Introduction to the inner Solar System**

The endeavor of planetary science is built upon a vast history of observations for the terrestrial planets within our Solar System (e.g., de Pater & Lissaur 2015; Horner et al. 2020, and references therein). Our inventory of terrestrial planets - Mercury, Venus, Earth, and Mars - provide us with natural laboratories from which to investigate crucial questions regarding the circumstances under which temperate surface conditions may arise and be sustained. To facilitate these studies, robotic explorations of the Solar System has supplied direct measurements of terrestrial planetary atmospheres and surface environments, including flybys of and subsequent landings on both Venus and Mars (e.g., Fjeldbo et al. 1966; Neugebauer & Snyder 1966; Avduevskij et al. 1971; Keldysh 1977; Hess et al. 1977; Toulmin et al. 1977). Such measurements form the foundation of fundamental models that describe atmospheric and geological planetary processes, as well as the origin and evolution of planetary systems (e.g., Goldreich & Soter 1966; Guillot 1999; Lodders 2003; Kane et al. 2021).

Although much has been learned regarding the present atmospheric and surface states of the terrestrial planets, many outstanding questions remain regarding their origins, evolution, and relationship to their interiors. Shown in Figure 1 are schematic cross-sections for the four Solar System terrestrial planets, that demonstrates their diversity of sizes, atmospheres, and interior structures. For simplicity, the details of crustal composition and mantle structure are not shown for Earth. Note that there is still considerable uncertainty regarding the state of the cores of Venus and Mars, particularly whether there are liquid and solid components, as for Earth's core.

The flanking of Earth by Venus and Mars has created a paradigm whereby we may be witnessing end-member states of planetary habitability (e.g., present Venus being too hot and Mars being too cold for sustaining conditions suitable for surface liquid water), from which crucial lessons can be extracted regarding Earth's own evolution. As Earth's sibling planet, understanding the reasons for Venus' current hostile surface, especially if Venus passed through a habitable period, are crucial for modeling the evolution of large rocky planets generally, both in our Solar System and beyond (Kane et al. 2014; Way et al. 2016; Kane et al. 2019; Way & Del Genio 2020; Ostberg et al. 2023, Gillmann et al., 2024). On the other hand, Mars represents a case study for a world that had a potentially habitable period, but which has undergone major changes in internal and surface properties as the interior cooled and the atmosphere was lost (Brain et al. 2016; Ehlmann et al. 2016; Banerdt et al. 2020). Resolving the numerous remaining questions regarding the conditions and properties of our local inventory of rocky worlds is thus critical for informing planetary models that show how surface conditions can reach equilibrium states that are either temperate and habitable, or hostile with thick and/or eroded atmospheres.

**Section 2. Solar evolution and terrestrial planets**

Besides a planet's internal thermal evolution, solar input is an important force behind planetary climatic evolution. Radiation from the Sun directly affects the surface temperature and the vertical and lateral structure of atmospheres. Even though solar radiation interacts with different atmospheres in different ways, it is also, through climate and as a source of energy, critical to the evolution of habitability.

Before the main sequence, the Sun went through the protostar and T-tauri phases (lasting respectively ~ $10^5$ and $10^7$ yr; Montmerle et al. 2006). This time coincides with the existence and dissipation of the circumstellar disk (Haisch et al. 2001; Carpenter et al. 2005; Monsch et al., 2023) and the formation of the planets in the Solar System. During the T-tauri phase, stars are larger and thus have a higher luminosity, and are characterized by high radio and x-ray emissions and intense solar wind (Appenzeller 1989), which probably affected protoplanets and young planets from the very beginning including, for example, stripping of primary atmospheres.

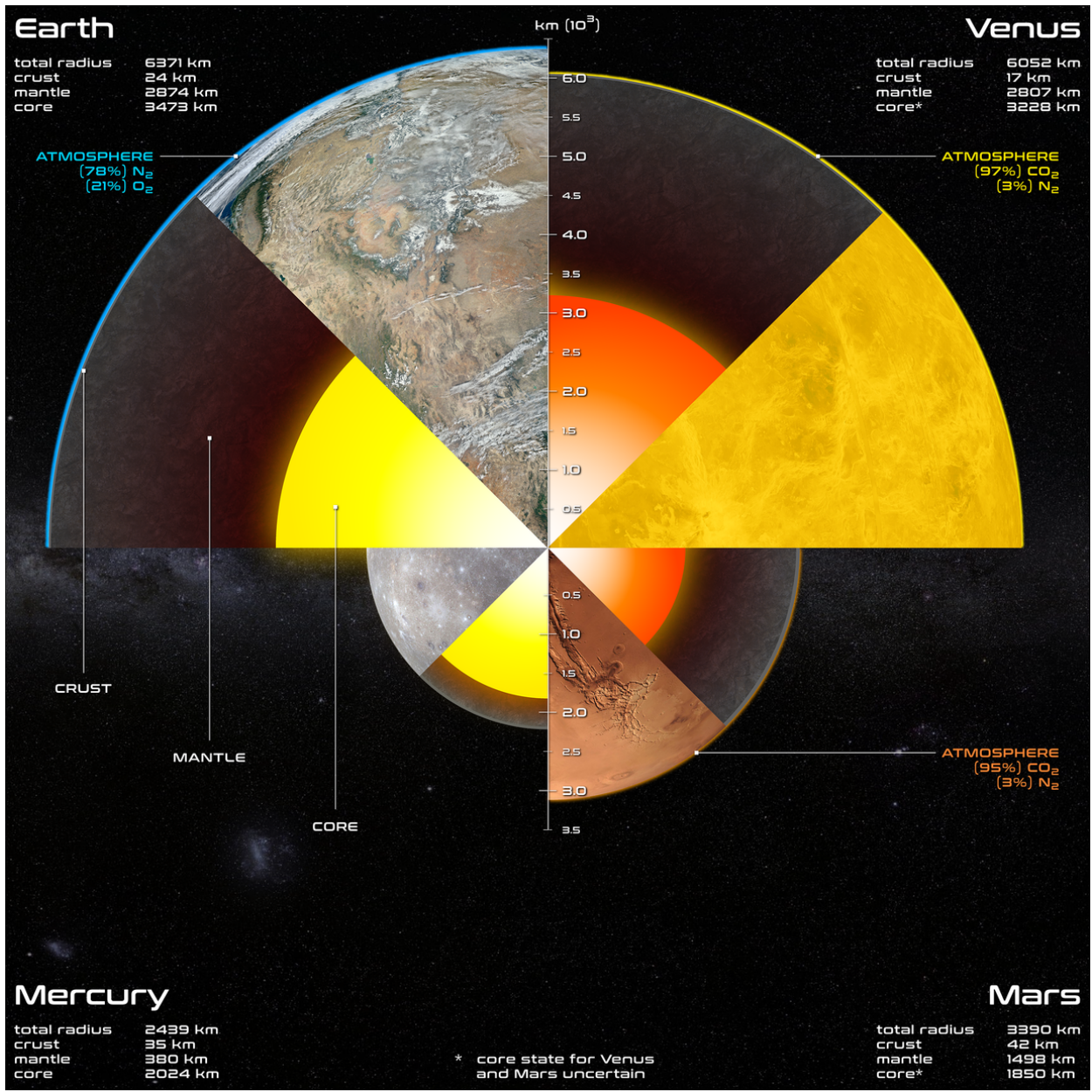

Figure 1: Schematic cross sections of the four inner Solar System planets, showing the major internal components (crust, mantle, and core) and atmospheric components. All cross sections are to scale. Credit: Kane et al. (2021).

After ~$10^7$ yr, the Sun entered the main sequence (Montmerle et al. 2006), where it exists today and likely for another ~5 Gyr. The radiation spectrum of the Sun is highly dependent on wavelength and intensity peaks around visible light. The Sun is currently a medium cool star of mid-G class with a surface temperature of about 5777 K. Since the onset of its main sequence, its temperature has increased from 70% of the present value. Throughout the last 4 Gyr of Earth's evolution, the Sun has helped maintain surface liquid water (See the Faint Young Sun Paradox, Sagan & Chyba 1997).

Planetary climate is highly dependent on the balance between absorbed solar radiation and the outgoing longwave radiation emitted by the surface and atmosphere (primarily IR), the latter of which is crucially dependent on the composition of the atmosphere (Pierrehumbert 2010).

The Sun also produces a magnetic field through an internal dynamo (Solanki et al. 2006). That magnetic field governs the outer structure of the Sun, including solar wind, solar flares and coronal structure and variations. In turn, the solar wind and flares interact with planetary atmospheres and magnetic fields.

Solar radiation can affect planets through another mechanism: atmospheric escape (see Gronoff et al. 2020 for a full discussion). Two types of escape processes are generally considered. The first is thermal loss, in which atoms gain enough energy to exceed escape velocity (either in the tail end of the energy distribution, termed "Jeans escape", or in bulk, for light atoms heated by intense EUV or by a magma ocean, called "hydrodynamic escape" (Jeans 1955; Chamberlain 1963; Hunten 1973, Kasting & Pollack 1983). Hydrodynamic escape is especially relevant for hydrogen atmospheres, small bodies and intense radiation/heating, and therefore to numerous inner planets in the early Solar System.

The second escape mechanism, non-thermal escape, covers everything else, but especially those mechanisms that are dependent on the interaction of the upper atmosphere with radiation from the Sun (e.g. ions being accelerated by the solar wind, electrical fields or along magnetic lines or through photochemical reactions. See Lammer 2013; Lammer et al. 2018, Gillmann et al. 2022). Non-thermal escape has been found to be comparatively inefficient, under Venus present-day conditions (for example, Persson et al. 2020) and could have had a small influence on atmosphere/surface conditions, although one should keep in mind that the past composition and structure of the atmosphere also govern escape and yet are are major unresolved factors (Gillmann et al. 2022). At present, interestingly enough, loss rate estimates for Earth, Mars and Venus are roughly similar (Ramstad & Barabash 2021) in spite of the differences in atmosphere composition, state and the presence/nature of their respective magnetic environments. Further, the question of the importance of the magnetic field to shield a planet from non-thermal escape is currently debated (Tarduno et al. 2014; Gunnell et al. 2018; Gronoff et al. 2020; Gillmann et al. 2024).

The 11-year solar cycle (the periodic variation of the number of sunspots, irradiance, short-wavelength radiation, flare index, etc. Coles et al. 1980; Lean 1997) is caused by the 22-year magnetic reversal cycle of the Sun. In turn, this cycle affects planetary climate (Haigh 1999; Gray et al. 2010). It also affects escape and can help us estimate loss rates at various EUV incident fluxes (Chassefiere et al. 2007; Gillmann 2009b, 2010, Persson et al. 2020).

On the longer timescales, the solar magnetic field is likely to have varied during the history of the Solar System. It is largely accepted that the Sun's magnetic activity was much higher during its early evolution, as the star spun faster than at present, in particular before it entered the main sequence (for a full review, see Gudel et al. 2007). Therefore, solar wind, UV, extreme UV and X-ray fluxes (unlike luminosity) would have decreased with time. "How much?" is a more complicated question. A comprehensive study of Sun-like stars indicates that EUV to X-ray emissions from the early Sun could exceed present levels by orders of magnitude (Ribas et al. 2005). However, the intensity of the emission depends on the rotation rates of the stars, which are highly variable (from observation of stars in <500 Myr old stellar clusters). "Fast rotating" young stars can reach about a hundred times the rotation rate of the modern Sun, whereas moderate rotators could rotate at about ten times that rate and slow rotators would reach a few times the Sun's rotation rate (Tu et al., 2015). Corresponding luminosities for young stars in the EUV and X-ray range of the spectrum can reach on the order of a 1000 times present mean solar luminosity at similar wavelengths for fast rotators, a few 100 times for moderate rotators and between a few tens to a hundred times the reference for slow rotators.

Not only is the luminosity history of the Sun uncertain but the length of the time the star spends in this early intensely radiating state varies with rotation rate, only then converging to a unique mass-dependent value (Soderblom et al. 1993). Therefore, these large short-wavelength luminosities could affect young planets for possibly hundreds of million years (~600 Myr, Gondoin, 2017), with critical consequences for their early atmospheres. Because of this convergence in the star's later evolution, it is not possible to estimate the initial rotation rate from present stellar observation only.

The precise rotational rate of the Sun in the past is uncertain. Only 70% of the solar mass stars studied by Johnstone et al. (2015) are slow and moderate rotators. However, observation of the depletion of sodium and potassium in lunar samples led Saxena et al. (2019) to propose that the Sun would have been a slow rotator, and modeling of atmospheric escape suggested that a rapidly rotating Sun would lead to unreasonably high amounts of H from the nebula being necessary in the primitive atmospheres of early planets in the Solar System (e.g. Odert et al., 2018). After the first 600 Myr, the evolution of the solar flux is comparatively better understood (e.g. Ribas et al. 2005).

In summary, the Sun has played a significant role in the climate evolution of the Solar System terrestrial planets and their potentially habitable surface conditions through time, largely through simultaneous contributions to their energy budgets and atmospheric loss processes. However, there remain uncertainties in the temperature and rotational evolution of the Sun which, in turn, propagate to uncertainties in solar EUV and X-ray emission that would have greatly impacted the early atmospheres of the terrestrial planets. These same issues are an increasingly important aspect in characterizing terrestrial exoplanets and their atmospheric states for a variety of spectral type host stars (Lammer et al. 2003; Roettenbacher & Kane 2017; Kane et al. 2020; Kite & Barnett 2020). Studying the Sun is thus a crucial template for not only our own Solar System planets, but planetary evolution in general.

**Section 3. The Evolution of Venus: Lessons learned from Earth's Twin**

Venus is often seen as Earth's planetary sibling because we do not know any other planet that has more in common with Earth. Venus has even been called a window into Earth's distant past (Hansen, 2007, 2018) or future (Way et al., 2016), and a glimpse of possible exoplanetary bodies (Kane et al. 2014; Kane et al. 2019; Way et al., 2023). However, a cursory survey of Venus' surface and atmosphere is enough to convince oneself that the planets certainly have not shared a similar fate: with a present surface temperature and pressure of ~740 K and 92 bar, respectively, Venus is far from an Earth-like habitable environment.

Unfortunately, the distant past of Venus remains hidden from us. One peculiar feature of Venus' surface, its apparently young and uniform age (250–1000 Myr old, Schaber et al., 1992, McKinnon et al., 1997), hinders attempts to uncover its past evolution (in contrast with Mars: see Section 5). Here, we gather the indirect evidence relevant to the formulation of plausible evolution scenarios for Venus, and its divergence from Earth.

Venus' orbit is the closest to Earth's of any of the other planets, and its atmosphere receives comparable amounts of energy from the Sun. Similar mechanisms likely presided over its formation, and its building blocks could mostly come from the same sources (McCubbin & Barnes, 2019, Greenwood et al., 2018, Izidoro et al., 2022). The similar size, mass, and thus density of the two planets further support the likelihood of a similar (but not fully identical) bulk composition and structure (Kaula 1994, Basilevsky and Head, 2003, Shah et al., 2022) despite large remaining uncertainties in the accretional and formational history of Venus (hampered further by the lack of known Venus samples).

Approximately 80% of the surface of Venus is covered with relatively young basaltic plains broadly comparable to material that could be found on Earth. The planet also exhibits numerous volcanic features (Ghail et al., 2024). In the absence of plate tectonics, the widespread basaltic magmatism is by definition intraplate and analogies have been sought in Earth's intraplate magmatic record. The largest basaltic events have been considered analogous to large igneous provinces (LIPs) (e.g., Head and Coffin, 1997; Hansen 2007; Ernst, 2014; Buchan and Ernst, 2021; El Bilali et al. 2023, El Bilali and Ernst, 2024). Special features on Venus, termed coronae, are manifest as circular volcano-tectonic structures featuring abundant fractures and concentrated along rifts, and in part may be analogous to terrestrial mantle upwelling-generated magmatism associated with giant circumferential dike swarms (e.g., Buchan and Ernst, 2021). The study of coronae has provided a valuable source of geophysical constraints for comparison between Venus and the Earth, including estimates for heat fluxes and lithospheric and elastic thicknesses, indicating that active regions of Venus may be more similar to Earth than previously suggested (Smrekar et al., 2023). Several formation mechanisms have been proposed, involving mantle upwellings, lithospheric dripping, and other processes (Stofan et al., 1991, Gerya, 2014, Gulcher et al., 2020, Tackley and Stevenson, 1991, Smrekar et al., 2023). In particular, the plume-induced roll-back subduction scenario is theorized to be a possible analog for subduction initiation on early Earth leading to the onset of plate tectonics (Gerya et al., 2015, Davaille et al., 2017).

About 8% of the Venus surface is covered by deformed highlands, the so-called tesserae (Ivanov and Head, 2011, Whitten et al., 2021). Tessera-covered highlands have been suggested to be analogs to Earth's continents (e.g. Nikolaeva et al., 1988, Hashimoto et al., 2008, Romeo and Capote, 2011), although this interpretation has been challenged (e.g., Karlsson et al., 2020). Additionally, tesserae have yet to be fully characterized, and other scenarios for their formation have been proposed (Hansen, 2006). At least some tesserae have further been suggested to be evolved igneous (felsic) rocks, which could require surface water, if formed by similar processes as on Earth (Shellnutt, 2013), although this formational mechanism, too, is disputed (Nimmo and Mackwell, 2023).

In any case, the planet has clearly been volcanically and tectonically active in the geologically recent past, which is further confirmed by evidence of present activity (e.g. Smrekar 1994, Smrekar et al., 2010, 2023, Herrick and Hensley, 2023, Gulcher et al., 2020). Surface ages have been suggested to be the consequence of either catastrophic volcanic resurfacing events or by ongoing "patchy" volcanism (Strom et al., 1994, Turcotte et al., 1999, Bjonnes et al., 2012, Herrick et al., 2023, Rolf et al, 2022). Localized rifting, deformation, or corona-related subduction on Venus differs from Mars' more static, conventionally "stagnant lid" tectonic regime (Rolf et al, 2022 and references therein). It also differs from Earth's modern plate tectonic regime, as Venus lacks a well-defined global network of plates. Other tectonic regimes have been suggested for Venus instead: an episodic-lid regime (alternating between active- and stagnant-lid tectonics marked by lithosphere overturns separated by static quiet periods, Turcotte 1993, 1995, Armann and Tackley 2012), a deformable episodic lid (episodic overturns, but with constant high surface mobility and non-localized deformation, due to a weak lithosphere; Tian et al., 2023), or a plutonic squishy lid (most melt is not extruded, forming a fragmented warm ductile lithosphere, with overall low surface mobility but a rather high heat flux; Rozel et al., 2017, Lourenco et al., 2020).

Such regimes could be representative of early Earth, with a thinner lithosphere and warmer convecting mantle, before modern plate tectonics (Rozel et al., 2017). Venus could illustrate the transition from a squishy-lid regime to the onset of plate tectonics. On the other hand, it has also been suggested that the relatively quiescent modern Venus could illustrate the end of a mobile convection pattern and ongoing transition into the stagnant regime (Weller and Kiefer 2020, Rolf et al., 2022). It is highly likely that the planet's convection regime has changed over the history of Venus, like on Earth, possibly in conjunction with surface conditions (Gillmann and Tackley, 2014, Noack et al., 2012, Weller et al., 2023). However, it should also be noted that the role of mantle upwellings and diapirs (and the intraplate magmatism that they produced) may be consistent through Venusian history, based on the terrestrial record. On Earth it has been shown that the expression of mantle upwellings, as LIPs, is approximately constant back to the Archean, and the more recent compilation (Ernst et al. in Chapter 3 of this volume) shows that Archean LIP record (at least back to 3.5 Ga) is approximately as dense as the post-Archean LIP record. Together, these findings suggest that plume production was approximately constant for 3.5 Gyr of Earth history through possibly different convective regimes: plate tectonics, plutonic squishly-lid, etc.

Our inability so far to find a self-generated magnetic field on Venus (Russel et al., 1980) constitutes further evidence that the interiors of Venus and the Earth behave differently (Dumoulin et al., 2017). Three main possibilities exist for Venus (see O'Rourke et al., 2018). (i) The core is (nearly) fully solid (Stevenson et al., 1983), which would be surprising, because such a state would imply that heat extraction on Venus has been even more efficient than at Earth with plate tectonics. (ii) The core loses heat slowly, due to an insulated, hot mantle, possibly comparable to ancient Earth (van Thienen et al., 2005). (iii) The core is stably stratified (Jacobson et al., 2017) and resistant to convection, perhaps due to the way it was formed (no giant impact for instance, unlike Earth). Hypotheses (i) and (ii) imply differences in thermal evolution, whereas (iii) would point to a singular event at the planet's origins.

Venus' atmosphere is possibly the only remaining witness of its early evolution. It also displays the most striking difference between Earth and Venus. Venus' atmosphere boasts a surface pressure of about 92 bar (~4.8 x $10^{20}$ kg), with $CO_2$ (96.5% vol.), $N_2$ (about 3.5% vol.), and minor species ($SO_2$, Ar, $H_2O$, CO, He, Ne) (e.g., Fegley, 2014,

Lecuyer et al., 2000). The origins of the atmosphere of Venus and its evolution are highly debated (Gillmann et al., 2022, Salvador et al., 2023). Venus' $CO_2$ atmospheric content is thought to be similar to that of shallow reservoirs (including carbonate deposits) on Earth (Lecuyer et al., 2000) and the two planets' atmospheres contain comparable masses of $N_2$, suggesting that the onset or continuation of $CO_2$ recycling on the Earth could be a major cause for the divergence between the two planets (Gillmann et al., 2024). Although Venus' large $N_2$ inventory hints at efficient outgassing, radiogenic argon measurements indicate that Venus outgassed less than the Earth (Kaula, 1999, Avice et al., 2022). This apparent conflict might be resolved if outgassing occurred very early during the history of Venus, when radiogenic argon was not yet formed (Gillmann et al., 2022). Venus' atmospheric deuterium/hydrogen ratio is large compared to Earth, implying a strong fractionation (Donahue et al. 1982, Grinspoon, 1993). Together with the tiny amount of water in the atmosphere of Venus, this condition could be caused by long-term escape of H from an ancient water reservoir. However, it has proven challenging to quantify the amount of lost water (Grinspoon 1993; Gillmann et al., 2022: Avice et al., 2022).

Hints regarding the evolution of Venus and its water do not allow us to establish unequivocally a single scenario. However, two main end-member pathways have emerged to explain available observations (Figure 2, Gillmann et al., 2022). Initial conditions (hot, with water as vapor, versus cool, with liquid water) appear critical to define the evolution scenario the planet follows (Turbet et al., 2021).

(i) For scenario 1, stable conditions for liquid water may never have been reached, leading to a long-lived magma ocean and extended water loss (Hamano et al., 2013, Gillmann et al., 2009a). Venus' atmosphere dries up in the first few hundred million years, either because water is trapped in the solid planet or because it is outgassed and lost to space. Under this scenario, Venus would by then have developed a dense $CO_2$ atmosphere. The past ~4 Gyr of Venus' evolution would have consisted of moderate outgassing of $CO_2$ up to the present observed level and limited water outgassing/escape. If so, then the fate of the planets is decided by the magma ocean evolution (and is currently being investigated; Salvador et al., 2024, Bower et al., 2022), through a combination of the effects of insolation and outgassing as a function of interior composition (redox state, relative abundances of H, C, O).

(ii) On the other hand, if Earth and Venus shared similar initial conditions, Venus may have cooled down fast, leading to water condensation and a form of $CO_2$ sink. A thin Earth-like $N_2$ atmosphere would have survived for some time, along with possible habitable conditions (Way and del Genio, 2020). Given the compared $CO_2$ inventories of Earth and Venus, differences between the two planets could have emerged when Earth was able to regulate its climate through the long-term carbon cycle whereas, for some reason, Venus was not (Driscoll and Bercovici 2013), perhaps due to differences in mantle dynamics or more complex combinations of planetary characteristics, including insolation. Alternatively, after a habitable period (Westall et al., 2023), existing carbonate reservoirs were destabilized. The nature and timing of the transition out of the habitable/temperate Venus conditions remain unknown, although volcanic events, in the form of several concomitant LIP-like events, have been suggested (Way et al., 2022). The apparently stratigraphically oldest rocks on Venus, the geologically complex tesserae, have been proposed to preserve evidence of fluvial erosion (Khawja et al. 2020), and if correct the timing of the climate transition would be syn- or post-tesserae time.

It remains uncertain if Venus and Earth followed parallel evolutionary paths for a time. Recent models seem to favor an early divergence, during the magma ocean phase, or the first billion years, in order to satisfy modern atmosphere composition constraints (Gillmann and Tackley 2014, Gillmann et al., 2020, 2016, Warren et al., 2023, Weller et al., 2023).

Key measurements are missing to settle the question and form a definitive picture of ancient Venus. For example, investigation of the nature of tesserae could constrain the past presence of Water, and study of volatiles in volcanic plumes in the atmosphere could inform whether the interior of Venus is dry or not (Widemann et al., 2023). So much is not fully understood about Venus that it remains challenging to develop solid parallels with the Earth. However, understanding the divergence between the two planets is a key to terrestrial planet evolution and habitability, and its potential loss (Kane & Byrne 2024).

**Section 4. Hadean Earth evolution as a gateway to Archean habitability and life**

Any discussion about evolution of the Archean Earth would benefit from at least a cursory glance at the Hadean that came before and set the stage. This multi-disciplinary book devoted to the Archean is no exception (see also Jones et al, this volume). Perhaps most notable is how quickly Earth developed and then sustained, to present time, a habitable surface environment that included liquid water oceans (Carlson et al., 2014). The same could not be said for Earth's nearest neighbors closer (Venus) and farther from (Mars) the Sun. Most notably, Venus and Mars both lost appreciable surface water, assuming they once had it in the first place. The details of water for both are certainly less clear in terms of temporal and spatial distributions and persistence, particularly for Venus, although Mars' watery past becomes clearer and more convincing with each new mission. An understanding of Archean habitability, and the continuation of those conditions to today, is fundamental to characterizing our Solar

System's history and deducing the potential planetary evolutionary pathways in systems beyond our own.

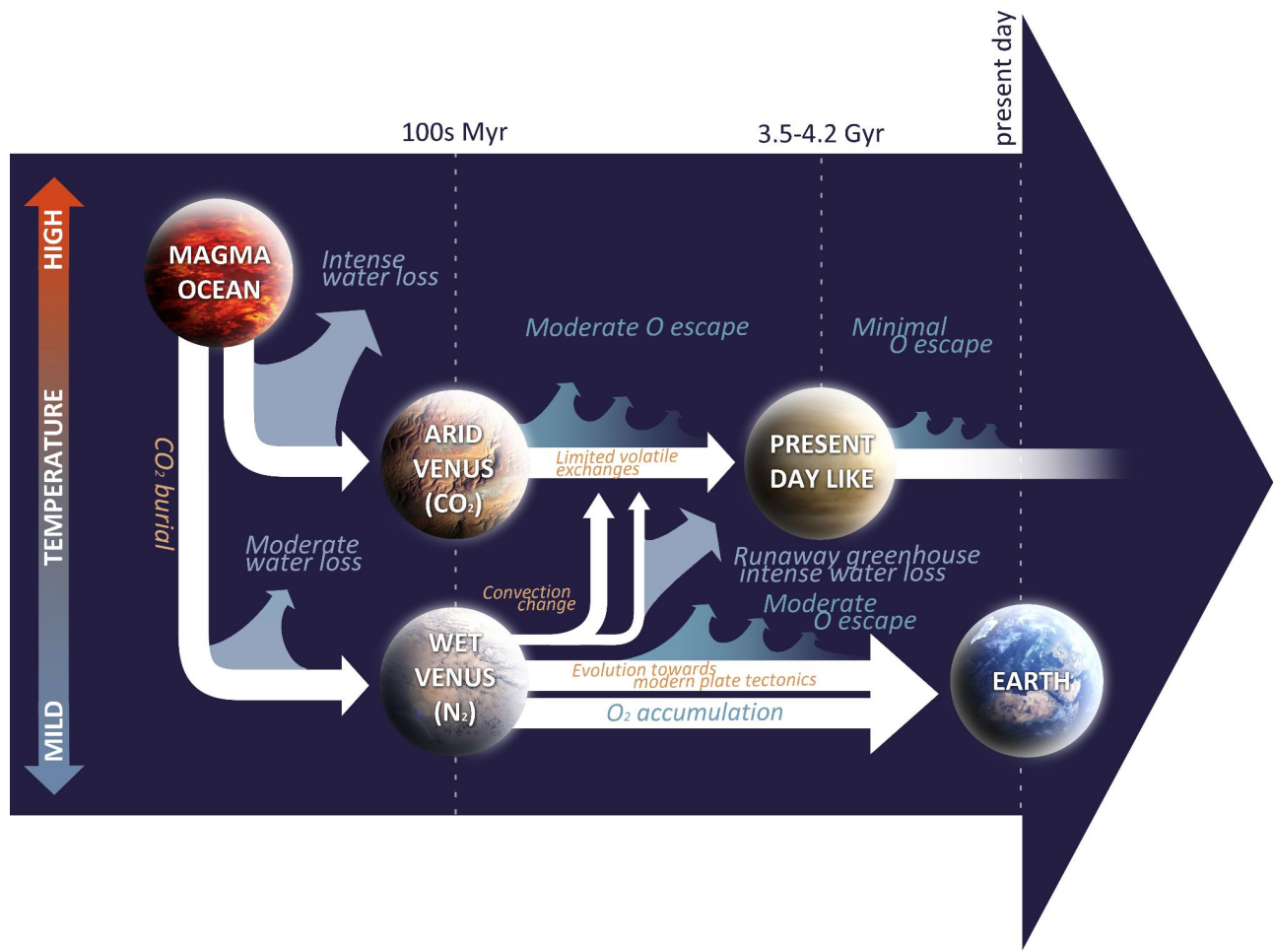

Figure 2: End-member scenarios for the evolution of Venus from its magma-ocean stage to the present. A first-order comparison to Earth is shown. White arrows indicate evolutionary pathways, pale blue arrows indicate escape processes where thermal escape is dominant, and deep blue arrows indicate non-thermal escape is dominant. Orange writing indicates interior or surface processes. (Adapted from Gillmann et al., 2022; Figure credit: C. Gillmann)

The many pieces regarding early oceans and the conditions for life's start on early Earth are far less conjectural (Fig. 3) and are discussed in detail elsewhere, including the contribution by Jones et al. in this volume. As such, we provide here only a quick overview of the most salient details, touching on the highlights as related in particular to the antecedent processes and products that led to an Archean Earth rich in life and habitable environments. The reader will be left with a sense that the Hadean Earth (>4.0 billion years ago, Ga) was not as hellish as the name implies and that the rich biosphere of the Archean is an expected outcome given the hundreds of millions of years of life-favoring conditions that came before.

The clock starts for Earth's habitability soon after the Moon-forming impact, which is placed within the first 10–100 million years of our history (Canup, 2004; Chambers, 2004). Liquid surface water and likely oceans followed soon after (Mojzsis et al., 2001), despite a faint young Sun that challenged the requirement for an early clement climate. The logical solution to warm conditions beneath a relatively cool Sun is to invoke greenhouse gases, in particular methane ($CH_4$) and carbon dioxide ($CO_2$). Methane, once popular in early climate models and origin-of-life scenarios (Miller and Urey, 1959), has mostly been taken off the Hadean table because of compelling evidence for early core formation (Kleine et al., 2009; Carlson et al., 2014). The removal of reducing power during core formation would have left the mantle too oxidizing to support appreciable abiotic $CH_4$ production (Trail et al., 2011). Biological production had to wait until the Archean at approx. 3.5 Ga (Wolfe and Fournier, 2018). The subject of subsequent mantle redox evolution is controversial (Kadoya et al., 2020a) and is covered elsewhere in the context of biospheric oxygenation during and following the Archean (e.g., Ostrander et al., this volume).

Biological production of $CH_4$ during the Archean must have ramped up its persistence and resulting concentrations, particularly in the absence of $O_2$ (See Goldblatt et al, this volume; Catling and Zahnle, 2020). The Hadean atmosphere would have been rich in $N_2$, but persistent atmospheric accumulation of $O_2$, resulting from oxygenic photosynthesis, did not occur until the Great Oxidation Event (GOE), which initiated at the end of the Archean and extended into the Paleoproterozoic (e.g., Ostrander et al., this volume).  Importantly, however, recent studies suggest the possibility of early production of reactive oxygen species such as $OH^-$ and $H_2O_2$ through entirely abiotic (i.e., geological) processes on mineral oxide surfaces (Stone et al., 2022; He et al., 2023). The presence of these oxygen species appears to have catalyzed early development of the enzymatic capability of dealing with damaging oxidation at the cellular level (Jabłońska and Tawfik, 2021). Early adaptations to oxidizing species primed life for the Archean and beyond, when $O_2$ accumulation began in the surface oceans and atmosphere (Lyons et al., 2024).

Very recently, however, Hadean $CH_4$ related to large and frequent impacts (Zahnle et al., 2020; Wogan et al., 2023) that stimulated transient production and accumulation has come back to center stage. These events might have episodically reset the atmosphere toward more reducing, $CH_4$-favoring conditions by delivering reduced compounds (e.g., metallic iron), thus compensating for the relatively oxidized mantle. The resulting transient reducing atmospheres would have facilitated abiotic synthesis of important prebiotic building block molecules critical to life's beginnings (summarized in Jones et al., this volume). Absent persistent Hadean $CH_4$, however, most of the burden of greenhouse warming would have fallen to $CO_2$ released abundantly from volcanoes, lower surface albedo effects in the absence of continents, and less cloud formation (Rosing et al., 2010), as well as potentially low sea ice cover (Olson et al., 2022). Another critical consideration is whether Hadean (and Archean) $CH_4$ rose to levels that led to haze formation in the $CO_2$-rich atmosphere (Arney et al., 2016). These hazes, analogous to those enveloping Saturn's moon Titan today, could have cooled instead of warmed the surface, perhaps offsetting some of the necessary greenhouse warming early in our history. On the other hand, organic hazes would also have shielded early prebiotic compounds and life from UV radiation, particularly in the absence of ozone ($O_3$) in the early $O_2$-free atmosphere. This possible haze benefit could have continued into and through the Archean (Arney et al., 2016).

Jones et al. (this volume) overview the possibilities for life's origins in that early world, including relationships to atmospheric composition, the role of hydrothermal systems, and possible solutions to the often-assumed requirement for wet–dry cycles for organic condensation reactions when continents were sparse or absent (reviewed in Korenaga, 2018) and likely submerged if present (Dong et al., 2021; Korenaga et al., 2021). High atmospheric $CO_2$ abundance would also have impacted the early oceans, most notably through correspondingly low (i.e., acidic) pH values (Halevy and Bachan, 2017), with important implications for how and where life first began (Jones et al., this volume).

It is remarkable that Earth seems to have become habitable within its first few hundreds of millions of years and maintained that state seemingly without interruption to the present. This persistence is a testimony to the power of feedbacks that drive and regulate climate evolution with notable success on Earth in comparison with our Solar System neighbors Venus and Mars. Most important is the silicate weathering negative feedback, whereby rising temperatures coinciding with a warming Sun are offset by increasing rates of $CO_2$ consumption through continental weathering, which is enhanced under warmer, wetter conditions. And we can thank diverse plate tectonic processes and their roles in such carbon cycling for this sustained thermostatic capacity over much of our history.

Early $CO_2$ abundance and persistence is also a theme that runs through discussions about the climates and habitability of Mars and Venus. We discuss the challenges Mars faced in retaining an atmosphere with the loss of its magnetic field (Section 5) and the failings of the silicate weathering feedback on Venus (Section 3). Earth has done very well by comparison, in part through the early development and persistence of a dynamo and resulting magnetic field that shielded against atmospheric loss via solar winds (Tarduno et al., 2014; Weiss et al., 2018). Earth experienced additional processes that modulated atmospheric $CO_2$ abundance even before large continents rose above the oceans. Those controls include $CO_2$ returned to the atmosphere through reverse weathering (clay formation) on the seafloor (Isson and Planavsky, 2018) and enhanced $CO_2$ uptake as fresh ejecta from large and frequent Hadean impacts was weathered (Kadoya et al., 2020b), all balanced against a very different carbon cycle and ocean chemistry (relative to post-Hadean times) in the absence of modern-style plate tectonics (Stern, 2018; Cawood et al., 2022) and biological carbon uptake and release. Apparently, this combination struck the right chord during the Hadean when continental weathering was likely minimal or absent, and high $CO_2$ would have been particularly beneficial given challenges to atmospheric methane accumulation and low solar luminosity.

On early Earth, $CO_2$ could have contributed to atmospheric reactions that, for example, led to production of $CH_4$ and subsequent organic molecules that may have fueled steps in the origins of life. Moreover, we can imagine solid-Earth processes that also impacted the beginnings of life and its subsequent evolution during the Archean. For example, our early interior was hotter than today, with higher degrees of partial melting that resulted in nickel (Ni)-rich ultramafic (komatiitic) igneous rocks (Mole et al., 2014) that gave way via later cooling to Ni-poor, phosphorus (P)-rich mafic (basaltic) lithologies (Johnson et al., 2014). The related composition, and alteration, of the early seafloor would have impacted the trace element inventories of Hadean oceans, affecting which elements were available to catalyze prebiotic reactions and would later be essential as enzymatic cofactors. Weathering of the Hadean seafloor, particularly if under relatively low pH conditions (Halevy and Bachan, 2017; Krissansen-Totton, et al., 2018), could have been a source of P (Syverson et al., 2021; Walton et al., 2023), a nutrient essential for the phosphorylation required to convert nucleosides to nucleotides in nucleic acids. During earlier chapters of Earth history, the recycling of P via the uplift and weathering of seafloors and continents rising above the oceans through plate tectonics likely played a vital role in (re)supplying P to the oceans.

Looking ahead along the timeline, with mantle cooling and corresponding decreases in the degree of mantle partial melting, Ni-rich rocks that fueled the enzymatic requirements of Archean microbial methanogens eventually transitioned to Ni-poor rocks that may have challenged methanogenesis (Konhauser et al., 2015). At the same time, the emergence of P-rich mafic rocks, including large igneous provinces exposed on the seafloor and continents (Cox et al., 2018; Chen et al., 2022), supported life in Archean and later oceans, despite nutrient limitations attributable to P scavenging (Bjerrum and Canfield, 2002; Reinhard et al., 2017; Rego et al., 2023) in iron-rich (ferruginous) anoxic waters. Ferruginous conditions continued through the Archean, Proterozoic, and into the Paleozoic (e.g., Poulton and Canfield, 2011, Planavsky et al., 2011). It is clear that limited or absent operation of modern style plate tectonics during the Hadean (Stern, 2018; Korenaga, 2021) set a very specific tone for the compositions of Earth's oceans and atmosphere, pathways of recycling, and possible scenarios for the beginnings of life and its early evolution. This situation all changed dramatically during the Archean, which featured continental growth, the emergence of land above the ocean surface, and supercontinent cycles (Korenaga, 2018).

A theme running throughout this Hadean narrative is the contribution from impacts that might have bolstered the beginnings of early habitability and life, and we could add delivery of reactive P to that list (Pasek, 2008; see also Jones et al., this volume). In stark contrast, however, is the long-held view that impacts at the end of the Hadean and early Archean were a massive roadblock to early life during the so-called Late Heavy Bombardment (LHB, 4.1 to 3.9 Ga) based on dynamical models and observations of cratering on the Moon and Mercury (Marchi et al., 2009, 2013; Bottke and Norman, 2017). Those data were ascribed to a purported resurgence in delivery of impactors, perhaps with deleterious consequence on Earth to the point of surface sterilization (Grimm and Marchi, 2014). One solution proposed for the survival of life on Earth invokes subsurface refugia that permitted shielding from the Hadean into the Archean (Abramov and Mojzsis, 2009).

More recently, however, the concept of an LHB and specifically a period of impacts that would have threatened life's persistence have been broadly challenged (e.g., Boehnke and Harrison, 2016; Cartwright et al., 2022). In the absence of an LHB, it is easier to imagine how metabolisms that included biological carbon fixation might have evolved by 3.7 Ga (Rosing and Frei, 2004) or even 4.1 Ga (Bell et al., 2015), and that the last universal common ancestor (LUCA) could have appeared contemporaneous with or even before the previously proposed LHB, even if the age of LUCA remains very poorly constrained. From this foundation of rapidly developed and persistently maintained habitability through the Hadean, microbial life advanced through the Archean via a cause-and-effect co-evolution with the environment, as witnessed by a wide diversity of metabolisms that drove environmental change. Principal among these changes are the beginnings of methanogenesis and oxygenic photosynthesis at perhaps 3.5 and 3.0 Ga, respectively, although these dates are not well determined and there are broad ranges discussed in the literature (Wolfe and Fournier, 2018; Planavsky et al., 2021). On the flip side, life also responded to favorable environmental change as recorded in the geologic record. One example of such a response is the dramatically increased geochemical footprints of organisms that used and produced oxidized chemical species in phase with the early stages of the GOE (Lyons et al., 2024). In many ways, the bounty of life and its favorable environments in the Archean should be no surprise given what came before. Remarkably, the complex interweaving of life–environment interactions and geological evolution benefited from and help set the course for persistent habitability on Earth. Any possibility of life on Venus and Mars suffered by comparison under the weight of collapsing habitability.

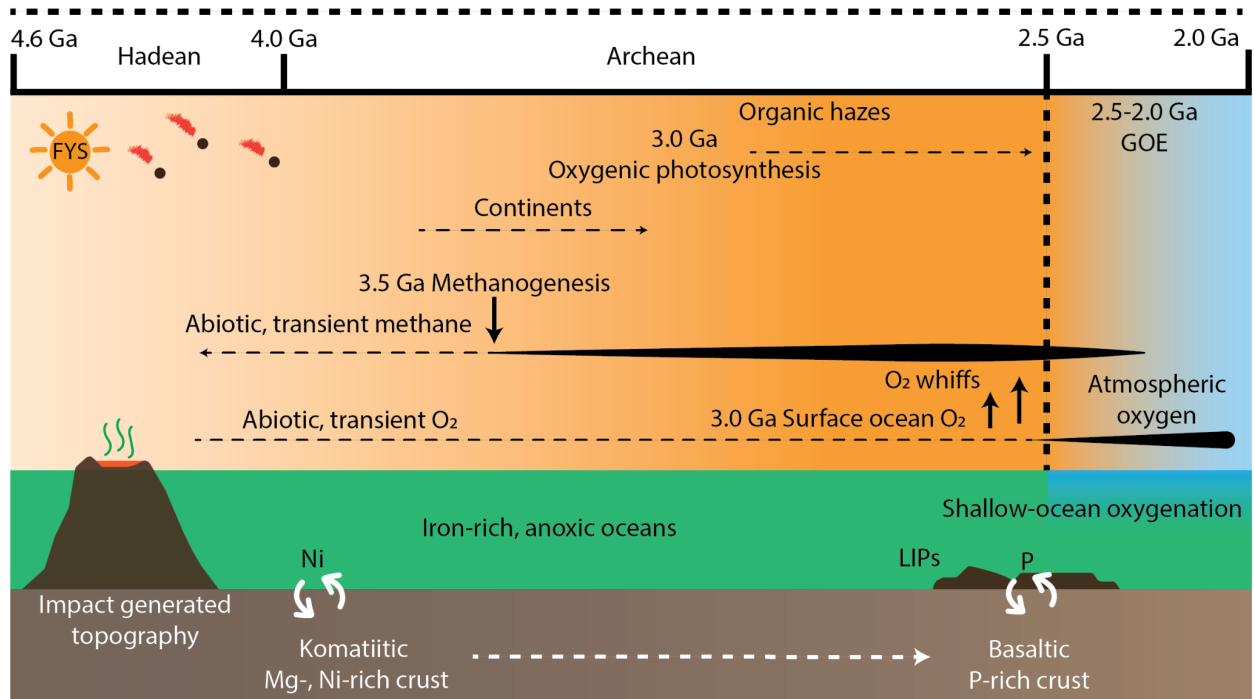

Figure 3: Illustration of potential atmospheric, oceanic, and crustal conditions, compositions, processes, and evolution on the early Earth. Approximate timing for the evolution of methanogenesis (3.5 Ga) and oxygenic photosynthesis (3.0 Ga), the emergence of continents, and crustal composition changes are given, although these timings are debated in the literature. There are various published estimates for the timing of the GOE. Here, a range of 2.5–2.0 Ga is given to encompass the potential extent of the transition to an oxygenated atmosphere. Abbreviations are as follows: Mg (magnesium), Ni (nickel), P (phosphorus), LIP (Large Igneous Province), GOE (Great Oxidation Event), and $O_2$ (oxygen).

**Section 5. Transient habitable surface conditions on ancient Mars**

Mars' predominantly ancient surface reveals an early collapse of habitability, underscoring the sensitivity of planetary-scale feedbacks that are necessary for the long-term maintenance of a habitable planet. Although Mars' lowest latitudes can currently experience above-freezing temperatures in the daytime, this does little to counter the planet's inability to convert its present-day ~20-30m global equivalent layer (GEL) of unbound water ice into sustained surface liquid water (Carr & Head, 2019). A non-existent global magnetic field and a sub-10 millibar atmosphere composed primarily of $CO_2$ give way to ionization radiation, while vanishingly low volcanic activity offers little hope for the consistent replenishment of atmospheric gases and the greenhouse warmth or biologically useful molecules they could provide.

Examinations of Mars' surface and atmosphere decisively indicate more moderate conditions in the distant past, possibly episodically (e.g., Mangold et al., 2016). With respect to habitability, most key planetary events appear to have occurred within 1.5 billion years of crust formation at no later than 4.547 Ga (Bouvier et al., 2018). Noachian (4.1-~3.5 Ga) and Hesperian (~3.5Ga-~3.0 Ga) terrains evidence a variety of mineralogical and sedimentological features best explained by surface or near-subsurface liquid water. These include valleys cut by streams, lacustrine sedimentary rocks within crater basins, and a variety of geochemical and/or mineralogical indicators of aqueous alteration including phyllosilicate formation (Poulet et al., 2005; Bishop et al., 2008; Fassett & Head, 2008a; Carr & Head, 2010; Grotzinger et al., 2014). Estimates for the exchangeable surface water reservoir during the Noachian range ~50-1500m GEL, with higher values resulting from an emphasis on crustal hydration as a water sequestration mechanism (Scheller et al., 2021; Jakosky & Hallis, 2024). Broadly, the warmth for liquid water is thought to have been generated similarly to the Archean Earth (4.0-2.5 Ga): a thick greenhouse atmosphere, mainly from volcanism, serving to retain the limited Solar energy provided by a less luminous young Sun (Sagan & Mullen, 1972; Catling & Zahnle, 2020). However, Mars' semimajor axis of 1.524 AU indicates it receives ~43% of the solar energy that Earth does (Wordsworth, 2016), meaning $CO_2$-$H_2O$ induced warming is inadequate and that high volumes of reduced gases such as $H_2$ and $CH_4$ may have been necessary to warm the surface (Ramirez et al., 2014; Wordsworth et al., 2021).

Estimates for Mars' initial inventories of $CO_2$ and $H_2O$ range 6-15 bar and 0.6-8.6 km GEL, respectively (Lunine et al., 2003, Brasser, 2012, Haberle et al., 2017). Much of this was lost prior to the formation of observable terrains (Cassata et al., 2022)—this is broadly supported by the isotopes of H, N, Ar, and Xe, although contradictions exist within these data (Haberle et al. 2017). Before ~4.1 Ga (pre-Noachian), the largest drivers of atmospheric loss were impact erosion, Jeans escape, and a possible early episode of hydrodynamic escape (Dreibus & Wänke, 1987; Melosh & Vickery, 1989). All of this occurred within the context of the young Sun's high extreme ultraviolet flux (EUV) (Tian et al. 2009). The collapse of Mars' global magnetic field at some point between 4.1 and 3.9 Ga (with a late dynamo perhaps extending to 3.7 Ga; Mittelholz et al., 2020), would have increased the importance of sputtering and dissociative recombination as escape mechanisms (Luhmann et al., 1992; Jakosky & Jones, 1997; Haberle et al., 2017). By the Noachian, the resulting total thickness of the atmosphere may have only been ≤400 mbar (based on the $^{40}Ar/^{36}Ar$ ratios of the ALH84001 meteorite; Cassata et al., 2012), while crater statistics place the upper limit at ~1.9 bar at 4.0 Ga and ~1.5 bar at 3.8 Ga (Warren et al., 2019). Results from modeling work regarding the volatile exchange processes are generally in line with these conclusions but can produce much lower atmospheric volatile inventories at 4.0 Ga (Gillmann et al., 2011, Leblanc et al., 2012). Another difficult reality is the global carbonate mineral record (Ehlmann et al., 2008; Niles et al., 2013), which does not indicate the sequestration of a multi-bar $CO_2$ atmosphere (Hu et al., 2015). However, experimental work has shown that carbonate formation may be geochemically unfavorable outside of ultramafic terrains, even under elevated $pCO_2$ (1 bar; Gil-Lozano et al., 2024).

While arguments persist for sustained warm and semi-arid conditions on ancient Mars (>3.5 Ga; Ramirez et al., 2018), paleolake hydrology (e.g., sediment transport from rivers) indicates that intermittent wet episodes amidst a freezing backdrop can explain surface morphology (McKay et al., 2005; Kite et al., 2019). The earliest Mars (>3.9 Ga) had rivers but appears to have been arid overall, as there is evidence of clay formation and erosion with little indication that water overspilled crater rims (Matsubara et al., 2018; Kite et al., 2024). Around the Noachian-Hesperian boundary, the planet's observable wetness optimum occurred (Rossman et al., 2002; Orofino et al., 2018; Kite et al., 2024). Seas formed, and water flow paths, some over 1000km in length, connected networks of lakes (Rossman et al., 2002; Fassett & Head, 2008b). Importantly, it may have only taken the equivalent of ~$10^5$ – $10^6$ years of continuous sediment flow to generate the valley networks (Orofino et al., 2018). Moving into the Hesperian, global drying ensued as volcanism waned to very low levels, and while river- and lake-forming climates extended into the Amazonian (~3.0 Ga to today), they became increasingly scarce both temporally and spatially (Kite et al., 2024).

The wet intervals in the Late Noachian and Hesperian could have resulted from $H_2$—sourced from volcanism, impact events, aqueous geochemistry, or a combination thereof—interacting with a ≥1 bar $CO_2$ atmosphere (Ramirez et al., 2014; Wordsworth et al., 2021). Furthermore, modelled alternations of reducing and oxidizing conditions as a function of $H_2$ flux may clarify rover observations of manganese oxides (Lanza et al., 2014; Wordsworth et al., 2021). Other possibilities for temporary warming include the intense greenhouse effect of high-altitude water ice clouds, which may explain sporadic Amazonian lakes (Kite et al., 2021) or the consequences of bolide impacts (e.g., Segura et al., 2012), although this scenario has been disputed (Turbet et al., 2020). In any case, the topography of the surface is important, because at atmospheric pressures >0.5 bar, low-to-mid-latitude highlands can function as cold traps for $H_2O$ ice due to adiabatic cooling (i.e., the icy highlands scenario; Wordsworth et al., 2013; Fastook et al., 2012). The location of most ancient valley networks broadly agrees with transient melt water flowing into topographic lows from highlands. Furthermore, Mars' chaotic, rapidly shifting, and wide-ranging obliquity (at times >40°) must also be considered (Laskar et al., 2004; Holo et al., 2018). Not only does extreme obliquity shift the latitudes of water formation, the range of possible $CO_2$-atmosphere collapse scenarios is dependent on it (Soto et al., 2015).

Throughout Mars' history, the subsurface may have been the most persistently habitable location (Ehlmann et al., 2011). Salinity and heat advection can sustain groundwater flow (Mellon & Phillips, 2001; Grasby et al., 2014) when surface conditions are uninhabitable. For instance, diagenesis is a common feature at Gale crater, including evidence of a groundwater episode approximately 1 billion years after original deposition the altered rocks (Bridges et al., 2015; Martin et al., 2017). Present-day subsurface habitability is also a possibility. The detection of methane by the Curiosity rover (and the Mars Express Orbiter) has sparked debate over its relevance to potential subsurface life (Webster et al., 2015; Yung et al., 2018; Giuranna et al., 2019; Korablev et al., 2019). Additionally, seismic data from the InSight Lander may indicate 1 to 2 km GEL of liquid water is buried in the mid-crust (Wright et al., 2024), and it has been suggested that radiolysis of groundwater could provide the energy and chemistry needed for life (Tarnas et al., 2021).

Mars teaches us that geologically brief events that are highly specific to a combination of regional and planetary-scale processes can leave long-lasting remnants of habitability. Whether intermittently habitable surface conditions prohibit a *de novo* origin of life is difficult to constrain (Pearce et al., 2018). Importantly, there remains much to explore in the oldest terrains (>3.7 Ga). Mineralogical studies of the Noachian highlands from orbit have identified relatively thick (several meters) aqueous weathering sequences, which may imply persistent wetness on timescales of millions of years (Loizeau et al., 2010; Carter et al., 2015). Eventually, *in-situ* geochemical measurements at these sites, as well as investigations of the subsurface, could represent new frontiers for Mars research and improve the prospects of past and present habitability.

C., Brunet, C., Hipkin, V., Léveillé, R., Marchand, G., Sánchez, P. S., Favot, L., Cody, G., Flückiger, L., Lees, D., Nefian, A., Martin, M., Gailhanou, M., Westall, F., Israël, G., Agard, C., Baroukh, J., Donny, C., Gaboriaud, A., Guillemot, P., Lafaille, V., Lorigny, E., Paillet, A., Pérez, R., Saccoccio, M., Yana, C., Armiens-Aparicio, C., Rodríguez, J. C., Blázquez, I. C., Gómez, F. G., Gómez-Elvira, J., Hettrich, S., Malvitte, A. L., Jiménez, M. M., Martínez-Frías, J., Martín-Soler, J., Martín-Torres, F. J., Jurado, A. M., Mora-Sotomayor, L., Caro, G. M., López, S. N., Peinado-González, V., Pla-García, J., Manfredi, J. A. R., Romeral-Planelló, J. J., Fuentes, S. A. S., Martinez, E. S., Redondo, J. T., Urqui-O'Callaghan, R., Mier, M.-P. Z., Chipera, S., Lacour, J.-L., Mauchien, P., Sirven, J.-B., Manning, H., Fairén, A., Hayes, A., Joseph, J., Sullivan, R., Thomas, P., Dupont, A., Lundberg, A., Melikechi, N., Mezzacappa, A., DeMarines, J., Grinspoon, D., Reitz, G., Prats, B., Atlaskin, E., Genzer, M., Harri, A.-M., Haukka, H., Kahanpää, H., Kauhanen, J., Paton, M., Polkko, J., Schmidt, W., Siili, T., Fabre, C., Wray, J., Wilhelm, M. B., Poitrasson, F., Patel, K., Gorevan, S., Indyk, S., Paulsen, G., Bish, D., Gondet, B., Langevin, Y., Geffroy, C., Baratoux, D., Berger, G., Cros, A., d'Uston, C., Forni, O., Gasnault, O., Lasue, J., Lee, Q.-M., Meslin, P.-Y., Pallier, E., Parot, Y., Pinet, P., Schröder, S., Toplis, M., Lewin, É., Brunner, W., Heydari, E., Achilles, C., Sutter, B., Cabane, M., Coscia, D., Szopa, C., Robert, F., Sautter, V., Le Mouélic, S., Nachon, M., Buch, A., Stalport, F., Coll, P., François, P., Raulin, F., Teinturier, S., Cameron, J., Clegg, S., Cousin, A., DeLapp, D., Dingler, R., Jackson, R. S., Johnstone, S., Lanza, N., Little, C., Nelson, T., Williams, R. B., Jones, A., Kirkland, L., Baker, B., Cantor, B., Caplinger, M., Davis, S., Duston, B., Fay, D., Harker, D., Herrera, P., Jensen, E., Kennedy, M. R., Krezoski, G., Krysak, D., Lipkaman, L., McCartney, E., McNair, S., Nixon, B., Posiolova, L., Ravine, M., Salamon, A., Saper, L., Stoiber, K., Supulver, K., Van Beek, J., Van Beek, T., Zimdar, R., French, K. L., Iagnemma, K., Miller, K., Goesmann, F., Goetz, W., Hviid, S., Johnson, M., Lefavor, M., Lyness, E., Breves, E., Dyar, M. D., Fassett, C., Edwards, L., Haberle, R., Hoehler, T., Hollingsworth, J., Kahre, M., Keely, L., McKay, C., Bleacher, L., Brinckerhoff, W., Choi, D., Dworkin, J. P., Floyd, M., Freissinet, C., Garvin, J., Glavin, D., Harpold, D., Martin, D. K., McAdam, A., Pavlov, A., Raaen, E., Smith, M. D., Stern, J., Tan, F., Trainer, M., Posner, A., Voytek, M., Aubrey, A., Behar, A., Blaney, D., Brinza, D., Christensen, L., DeFlores, L., Feldman, J., Feldman, S., Flesch, G., Jun, I., Keymeulen, D., Mischna, M., Morookian, J. M., Pavri, B., Schoppers, M., Sengstacken, A., Simmonds, J. J., Spanovich, N., Juarez, M. de . la T., Webster, C. R., Yen, A., Archer, P. D., Cucinotta, F., Jones, J. H., Morris, R. V., Niles, P., Rampe, E., Nolan, T., Fisk, M., Radziemski, L., Barraclough, B., Bender, S., Berman, D., Dobrea, E. N., Tokar, R., Cleghorn, T., Huntress, W., Manhès, G., Hudgins, J., Olson, T., Stewart, N., Sarrazin, P., Vicenzi, E., Bullock, M., Ehresmann, B., Hamilton, V., Hassler, D., Peterson, J., Rafkin, S., Zeitlin, C.,

Belgacem, I., Bell, J. F., Bender, S., Benna, M., Bentz, J., Berger, J., Berger, T., Berlanga, G., Berman, D., Bish, D., Blacksberg, J., Blake, D. F., José Blanco, J., Blaney, Á. D., Blank, J., Blau, H., Bleacher, L., Boehm, E., Bonnet, J.-Y., Botta, O., Böttcher, S., Boucher, T., Bower, H., Boyd, N., Boynton, W., Braswell, S., Breves, E., Bridges, J. C., Bridges, N., Brinckerhoff, W., Brinza, D., Bristow, T., Brunet, C., Brunner, A., Brunner, W., Buch, A., Bullock, M., Burmeister, S., Burton, J., Buz, J., Cabane, M., Calef, F., Cameron, J., Campbell, J. L., Cantor, B., Caplinger, M., Clifton, C., Caride Rodríguez, J., Carmosino, M., Carrasco Blázquez, I., Cavanagh, P., Charpentier, A., Chipera, S., Choi, D., Christensen, L., Clark, B., Clegg, S., Cleghorn, T., Cloutis, E., Cody, G., Coll, P., Coman, E. I., Conrad, P., Coscia, D., Cousin, A., Cremers, D., Crisp, J. A., Cropper, K., Cros, A., Cucinotta, F., d'Uston, C., Davis, S., Day, M., Daydou, Y., DeFlores, L., Dehouck, E., Delapp, D., DeMarines, J., Dequaire, T., Des Marais, D., Desrousseaux, R., Dietrich, W., Dingler, R., Domagal-Goldman, S., Donny, C., Downs, R., Drake, D., Dromart, G., Dupont, A., Duston, B., Dworkin, J. P., Dyar, M. D., Edgar, L., Edgett, K., Edwards, C. S., Edwards, L., Edwards, P., Ehlmann, B., Ehresmann, B., Eigenbrode, J., Elliott, B., Elliott, H., Ewing, R., Fabre, C., Fairén, A., Fairén, A., Farley, K., Farmer, J., Fassett, C., Favot, L., Fay, D., Fedosov, F., Feldman, J., Fendrich, K., Fischer, E., Fisk, M., Fitzgibbon, M., Flesch, G., Floyd, M., Flückiger, L., Forni, O., Fox, V., Fraeman, A., Francis, R., François, P., Franz, H., Freissinet, C., French, K. L., Frydenvang, J., Garvin, J., Gasnault, O., Geffroy, C., Gellert, R., Genzer, M., Getty, S., Glavin, D., Godber, A., Goesmann, F., Goetz, W., Golovin, D., Gómez Gómez, F., Gómez-Elvira, J., Gondet, B., Gordon, S., Gorevan, S., Graham, H., Grant, J., Grinspoon, D., Grotzinger, J., Guillemot, P., Guo, J., Gupta, S., Guzewich, S., Haberle, R., Halleaux, D., Hallet, B., Hamilton, V., Hand, K., Hardgrove, C., Hardy, K., Harker, D., Harpold, D., Harri, A.-M., Harshman, K., Hassler, D., Haukka, H., Hayes, A., Herkenhoff, K., Herrera, P., Hettrich, S., Heydari, E., Hipkin, V., Hoehler, T., Hollingsworth, J., Hudgins, J., Huntress, W., Hurowitz, J., Hviid, S., Iagnemma, K., Indyk, S., Israël, G., Jackson, R. S., Jacob, S., Jakosky, B., Jean-Rigaud, L., Jensen, E., Kløvgaard Jensen, J., Johnson, J. R., Johnson, M., Johnstone, S., Jones, A., Jones, J. H., Joseph, J., Joulin, M., Jun, I., Kah, L. C., Kahanpää, H., Kahre, M., Kaplan, H., Karpushkina, N., Kashyap, S., Kauhanen, J., Keely, L., Kelley, S., Kempe, F., Kemppinen, O., Kennedy, M. R., Keymeulen, D., Kharytonov, A., Kim, M.-H., Kinch, K., King, P., Kirk, R., Kirkland, L., Kloos, J., Kocurek, G., Koefoed, A., Köhler, J., Kortmann, O., Kotrc, B., Kozyrev, A., Krau, J., Krezoski, . ß G., Kronyak, R., Krysak, D., Kuzmin, R., Lacour, J.-L., Lafaille, V., Langevin, Y., Lanza, N., Lapôtre, M., Larif, M.-F., Lasue, J., Le Deit, L., Le Mouélic, S., Lee, E. M., Lee, Q.-M., Lee, R., Lees, D., Lefavor, M., Lemmon, M., Lepinette, A., Lepore, M. K., Leshin, L., Léveillé, R., Lewin, É., Lewis, K., Li, S., Lichtenberg, K., Lipkaman, L., Lisov, D., Little, C., Litvak, M., Liu, L., Lohf, H.,

Lorigny, E., Lugmair, G., Lundberg, A., Lyness, E., Madsen, M. B., Magee, A., Mahaffy, P., Maki, J., Mäkinen, T., Malakhov, A., Malespin, C., Malin, M., Mangold, N., Manhes, G., Manning, H., Marchand, G., Marín Jiménez, M., Martín García, C., Martin, D. K., Martin, M., Martin, P., Martínez Martínez, G., Martínez-Frías, J., Martín-Sauceda, J., Martín-Soler, M. J., Martín-Torres, F. J., Mason, E., Matthews, T., Matthiä, D., Mauchien, P., Maurice, S., McAdam, A., McBride, M., McCartney, E., McConnochie, T., McCullough, E., McEwan, I., McKay, C., McLain, H., McLennan, S., McNair, S., Melikechi, N., Mendaza de Cal, T., Merikallio, S., Merritt, S., Meslin, P.-Y., Meyer, M., Mezzacappa, A., Milkovich, S., Millan, M., Miller, H., Miller, K., Milliken, R., Ming, D., Minitti, M., Mischna, M., Mitchell, J., Mitrofanov, I., Moersch, J., Mokrousov, M., Molina, A., Moore, J. C., Moores, J. E., Mora-Sotomayor, L., Moreno, G., Morookian, J. M., Morris, R. V., Morrison, S., Mousset, V., Mrigakshi, A., Mueller-Mellin, R., Muller, J.-P., Muñoz Caro, G., Nachon, M., Nastan, A., Navarro López, S., Navarro González, R., Nealson, K., Nefian, A., Nelson, T., Newcombe, M., Newman, C., Newsom, H., Nikiforov, S., Nikitczuk, M., Niles, P., Nixon, B., Noblet, A., Noe, E., Nolan, D. T., Oehler, D., Ollila, A., Olson, T., Orthen, T., Owen, T., Ozanne, M., de Pablo Hernández, M. Á., Pagel, H., Paillet, A., Pallier, E., Palucis, M., Parker, T., Parot, Y., Parra, A., Patel, K., Paton, M., Paulsen, G., Pavlov, A., Pavri, B., Peinado-González, V., Pepin, R., Peret, L., Pérez, R., Perrett, G., Peterson, J., Pilorget, C., Pinet, P., Pinnick, V., Pla-García, J., Plante, I., Poitrasson, F., Polkko, J., Popa, R., Posiolova, L., Posner, A., Pradler, I., Prats, B., Prokhorov, V., Raaen, E., Radziemski, L., Rafkin, S., Ramos, M., Rampe, E., Rapin, W., Raulin, F., Ravine, M., Reitz, G., Ren, J., Rennó, N., Rice, M., Richardson, M., Ritter, B., Rivera-Hernández, F., Robert, F., Robertson, K., Rodriguez Manfredi, J. A., José Romeral-Planelló, J., Rowland, S., Rubin, D., Saccoccio, M., Said, D., Salamon, A., Sanin, A., Sans Fuentes, S. A., Saper, L., Sarrazin, P., Sautter, V., Savijärvi, H., Schieber, J., Schmidt, M., Schmidt, W., Scholes, D., Schoppers, M., Schröder, S., Schwenzer, S. P., Sciascia Borlina, C., Scodary, A., Sebastián Martínez, E., Sengstacken, A., Shechet, J. G., Shterts, R., Siebach, K., Siili, T., Simmonds, J. J., Sirven, J.-B., Slavney, S., Sletten, R., Smith, M. D., Sobron Sanchez, P., Spanovich, N., Spray, J., Spring, J., Squyres, S., Stack, K., Stalport, F., Starr, R., Stein, A. S. T., Stern, J., Stewart, N., Stewart, W., Stipp, S. S. L., Stoiber, K., Stolper, E., Sucharski, R., Sullivan, R., Summons, R., Sumner, D. Y., Sun, V., Supulver, K., Sutter, B., Szopa, C., Tan, F., Tate, C., Teinturier, S., ten Kate, I. L., Thomas, A., Thomas, P., Thompson, L., Thuillier, F., Thulliez, E., Tokar, R., Toplis, M., de la Torre Juárez, M., Torres Redondo, J., Trainer, M., Treiman, A., Tretyakov, V., Ullán-Nieto, A., Urqui-O'Callaghan, R., Valentín-Serrano, P., Van Beek, J., Van Beek, T., VanBommel, S., Vaniman, D., Varenikov, A., Vasavada, A. R., Vasconcelos, P., de Vicente-Retortillo Rubalcaba, Á., Vicenzi, E., Vostrukhin, A., Voytek, M., Wadhwa, M., Ward, J.,